\documentclass[prd,preprint,showpacs]{revtex4} \usepackage{latexsym}
\usepackage{amsfonts} \usepackage{amsmath} 

\newcommand{\kt}[1]{\ensuremath{|#1\rangle}}
\newcommand{\br}[1]{\ensuremath {\langle #1|}}
\newcommand{\bk}[2]{\ensuremath {\langle #1|#2 \rangle}}

\newcommand{\wig}{\ensuremath p\,\chi [\ms\,j]}

\newcommand{\vp}{{\bf p}}

\newcommand{\mmp}{\mathsf{p}}

\newcommand{\mk}{\mathsf{k}}
\newcommand{\vP}{{\bf P}}

\newcommand{\ms}{\mathsf{s}}

\newcommand{\slg}{{\rm SL}(2, \mathbb{C})}
\newcommand{\su}{{\rm SU}(2)}
\newcommand{\ptr}{\tilde{\mathcal{P}}^\uparrow_+}
\newcommand{\pt}{\tilde{\mathcal{P}}}
\newcommand{\al}{\alpha}
\newcommand{\La}{\Lambda}

\begin{document}

\title{Clebsch-Gordan Coefficients for the Extended Quantum-Mechanical Poincar\'e Group and Angular Correlations of Decay Products}

\author{N.L.~Harshman}
\affiliation{Department of Computer Science, Audio Technology and Physics\\
4400 Massachusetts Ave., NW \\ American University\\ Washington, DC 20016-8058}
\author{N. Licata}
\affiliation{Department of Physics\\ Randall Laboratory\\
University of Michigan\\ Ann Arbor, MI 48109-1120}

\begin{abstract}

This paper describes Clebsch-Gordan coefficients (CGCs) for unitary irreducible representations (UIRs) of the extended quantum mechanical Poincar\'e group $\pt$.  `Extended' refers to the extension of the 10 parameter Lie group that is the Poincar\'e group by the discrete symmetries $C$, $P$, and $T$; `quantum mechanical' refers to the fact that we consider projective representations of the group.  The particular set of CGCs presented here are applicable to the problem of the reduction of the direct product of two massive, unitary irreducible representations (UIRs) of $\pt$ with positive energy to irreducible components.  Of the sixteen inequivalent representations of the discrete symmetries, the two standard representations with $U_C U_P = \pm 1$ are considered.  Also included in the analysis are additive internal quantum numbers specifying the superselection sector.  As an example, these CGCs are applied to the decay process of the $\Upsilon(4S)$ meson.

\end{abstract}


\maketitle

\section{Introduction}

The intent of this paper is to present properties of the Clebsch-Gordan coefficients (CGCs) for unitary irreducible representations (UIRs) of the extended Poincar\'e group. The direct product of UIRs of $\pt$, the covering group of the extended Poincar\'e group, is no longer irreducible, but can be decomposed into a direct sum (or tower) of UIRs of $\pt$ whose invariants are the center-of-mass energy squared $\ms$ and total angular momentum $j$, as well as other discrete degeneracy parameters $\eta$.
The CGCs are the transformation coefficients between the direct product basis of the two UIRs and the basis vectors in each of the UIRs in the direct sum decomposition.  The CGCs for $\pt$ give a fully relativistic justification for angular correlation results involving P- and C-conserving scattering and decays.  They explicitly incorporate the fact that two standard representations of the discrete symmetries chosen for fermions and bosons are combined consistently.

In relativistic scattering systems this technique is called relativistic partial wave analysis~\cite{weinberg}.  For example, two particle in-states and out-states (out-observables) are constructed as the direct product of two wave functions (or more commonly, the singular basis vectors) that belong to UIRs of $\pt$ with invariants corresponding to the particle mass squared, spin, intrinsic parity and charge parity, and charges.  If the interactions satisfy Poincar\'e symmetry principles, the S-matrix and transition amplitudes are diagonal in the invariants of the UIRs appearing in the direct sum decomposition of the direct product of the individual particle UIRs.  Important conventions for this procedure using the helicity basis for the UIRs of the restricted Poincar\'e group $\ptr$ were established by Jacob and Wick~\cite{jacobwick} and has been incorporated into the tensor formalism of Chung~\cite{chung}.  Other CGCs for $\ptr$ employ the Wigner momentum (or instant form) basis and rely on the spin orbit coupling method in analogy to the non-relativistic case~\cite{joos,macfarlane};  We use this coupling scheme because then the discrete degeneracy parameters $\eta$ label UIRs of $\pt$ of definite parity (and charge parity in the case of totally neutral compositions).  The fact that partial wave analysis for relativistic states leads to conclusions similar to those for non-relativistic states is because of the rotation subgroup of $\pt$.

Besides kinematic analysis, the CGCs are used in a variety of contexts.  Employing CGCs for the extended Poincar\'e group in the transversity basis is useful for analyzing CP-violation in the decay of neutral mesons~\cite{transversity}.  In bound states relativistic quark models for hadrons, CGCs have been used for a variety of bases~\cite{kandpoly,poly} in the Bakamijan-Thomas construction for relativistic Hamiltonian dynamics~\cite{bandt}, where extensive use is made of the fact that the interacting system must also have $\ptr$ symmetry~\cite{weinberg,dirac49}.  The CGCs also play an important role in the definition of the relativistic Gamow vectors~\cite{hani}, extensions of the Wigner velocity (or point form) basis kets to certain irreducible representations of the Poincar\'e semigroup used in the analysis of the scattering and decay of unstable particles~\cite{rgv,zboson}.  None of the works cited above establish which of the 16 inequivalent representations of the discrete symmetries for a given mass and spin are chosen; this work makes that extension and also incorporates the simplest case of additive internal quantum numbers labeling the superselection sectors.

Field theory typically works with the irreducible, but non-unitary, spinor representations of $\pt$.  While working with basis vectors of UIRs of $\pt$ is equivalent to working with the solutions of wave equations for non-interacting fields\cite{wigbarg}, it is better suited for some of the applications described above.  The seminal paper in this approach is the classic work of Winger~\cite{wigner39}.  We will use results on the UIRs of $\ptr$ found in \cite{ruhl, schaaf} and of $\pt$ described in \cite{wigner64,goldberg,scurek}.

This paper will be concerned with the UIRs of $\pt$ associated to positive mass representations with momentum in the forward light cone and for which two invariants are the mass-squared $\ms_i=m_i^2$ and the intrinsic spin $j_i$.  For a particular representation space $\Phi(\ms_i,j_i)$, a complete set of commuting observables (CSCO) can be chosen, two of which correspond to the invariants and act multiplicatively on every vector in the UIR.  For the other four, three are typically chosen to be ``momentum''-like and one is a component of the spin-operator.  We work with the CSCO $\{M^2, W^2, {\bf P}, S_3(P)\}$, whose action will be defined below.

The reduction of the direct product of two UIRs of $\pt$ has some mathematical similarities to the CGC reduction of the representations of the rotation group ${\rm SO}(3)$ or its quantum mechanical covering $\su$.  However, $\pt$ is not compact and the UIRs are infinite dimensional.  The spectrum of $\Phi(\ms,j)$ that appear in the direct product of two massive UIRs of $\pt$ runs from $(m_1 + m_2)^2 \leq \ms <\infty$ and $j\in\{j_0, j_0+1\}$, where $j_0=0$ if both particles are bosons or fermions and $j_0=1/2$ if the direct product is mixed.  However, a given $\Phi(\ms,j)$ may appear multiple times in the reduction of the direct product of $\Phi(\ms_1,j_1)\otimes\Phi(\ms_2,j_2)$; this degeneracy can by quantified by invariant parameters which we combine as the set $\eta$ and include as a label on the direct sum $\Phi(\ms,j)_\eta$.  The physical interpretation of these parameters depends on the coupling scheme used; in our case of spin-orbit coupling $\eta=\{l,s\}$ the orbital angular momentum and total spin.  These parameters are particularly important for understanding the transformation properties of the direct sum UIRs under the discrete symmetries.  A few other complications arising from internal quantum numbers and spin-statistics will be incorporated below.

Our results break into three categories.  For compositions of a particle and its antiparticle, equation (\ref{paptcgc}) gives the CGC of $\pt$, including the relations between $\eta$ and the intrinsic parity and charge parity of the composite.  For compositions of particles that are their own antiparticles, equation (\ref{npcgcpt}) gives the CGC of interest.  In all other cases the intrinsic charge parity of the composite system is not physically meaningful and (\ref{cpcgcpt}) can be used.

The outline for this paper is as follows: first, the transformation laws and various bases for $\ptr$ will be defined.  Next, the direct product of two UIRs of $\ptr$ will be considered and CGCs expressed.  Then, after a few comments about additive internal quantum numbers and identical particles, the transformation laws and various bases for $\pt$ will be defined the CGCs for the covering group of the full Poincar\'e group $\pt$ will be defined.  For easier reading and to highlight the important results, some details have been moved to appendices.  Finally we will give an example with the $\Upsilon(4S)$ meson, the center of much research on CP-violation, and some indications of further directions this work will take.

\section{UIRs of the restricted Poincar\'e group}

First we consider the UIRs of the restricted quantum-mechanical Poincar\'e group $\tilde{\mathcal{P}}^\uparrow_+$.  The qualifier `restricted' means we only consider proper and orthochronous transformations, i.e. at first we will not consider space or time reflections; representations of these transformations will be adjoined in a subsequent section.  By `quantum mechanical', we mean that we will consider 
projective representations of $\mathcal{P}^\uparrow_+$, i.e. representations of the proper Poincar\'e group up to a phase, which is equivalent to considering UIRs of the covering group $\ptr$~\cite{wigner39}.

The group $\tilde{\mathcal{P}}^\uparrow_+$ is isomorphic to the semidirect product ${\rm SL}(2, \mathbb{C})\times \mathbb{R}^4$ and we will denote these elements as $(\alpha, a)$, where $\alpha\in{\rm SL}(2, \mathbb{C})$ and $a\in\mathbb{R}^4$.  The group multiplication law is 
\begin{equation}
(\alpha', a')(\alpha, a) = (\alpha'\alpha, a' + \Lambda(\alpha') a),
\end{equation}
where $\Lambda(\alpha)\in{\rm SO}(1,3)$ is a well-known two-to-one homomorphism~\cite{ruhl}.  The subset ${\rm SU}(2)\subset{\rm SL}(2, \mathbb{C})$ corresponds to the rotation subgroup of transformations; $a\in\mathbb{R}^4$ is the translation subgroup.

The choice  for CSCO $\{M^2, W^2, {\bf P}, S_3(p)\}$ leads to the Wigner 3-momentum spin basis for the expansion of the representation space $\Phi(\ms,j)$.  See Appendix A for more details on this CSCO and the space $\phi\in\Phi(\ms,j)$.  Summarizing, if $\phi\in\Phi(\ms,j)$ are chosen to be elements of the Schwartz space of ``well-behaved'' functions of the momentum, then improper eigenvectors $\kt{\wig}$, or Dirac eigenkets, of this CSCO  are elements of $\Phi^\times(\ms,j)$, the linear topological dual of $\Phi(\ms,j)$ and have the following properties:
\begin{eqnarray}
M^2\kt{\wig} &=& \ms \kt{\wig} \nonumber \\
W^2\kt{\wig} &=& \ms j(j+1)\kt{\wig}\nonumber \\
{\bf P}\kt{\wig} &=& {\bf p}\kt{\wig}\nonumber \\
S_3(P)\kt{\wig} &=& \chi\kt{\wig}.
\end{eqnarray}
Then $j$ is interpreted as the intrinsic spin of the particle/representation, ${\bf p}$ is the spatial components of the momentum of the basis ket, and $\chi$ is the third component of the ket transformed to the rest frame $p_R = (\sqrt{\ms}, 0 ,0 ,0)$.

Choosing a relativistically invariant normalization,
\begin{equation}\label{wignorm}
\bk{p',\chi'[\ms,j]}{\wig} = 2 \ms E(\mmp)\delta^3({\bf p'} - {\bf p})\delta_{\chi'\chi},
\end{equation}
gives the following form to the expansion of a vector $\phi\in\Phi$ with invariant measure
\begin{equation}
\phi = \frac{1}{\ms}\sum_\chi \int \frac{d^3{\bf p}}{2 E(\mmp)} \kt{\wig}\bk{\wig}{\phi},
\end{equation}
where $p = (E(\mmp), \vp)$ and $\mmp^2 = \vp^2$.
The somewhat non-standard factor of $\ms$ in these equations is chosen so that the state vectors and kets do not carry dimensional units, which will ensure that CGC's introduced later are unitless, infinite dimensional matrices.

For $\phi\in\Phi(\ms,j)$ and $\kt{\wig}\in\Phi^\times(\ms,j)$, the Poincar\'e transformations then are represented in this infinite dimensional basis as
\begin{subequations}\label{wigtrans}
\begin{eqnarray}
U(\alpha, a)\phi_\chi( p) &=& \br{\wig}U(\alpha, a)\kt{\phi}\nonumber \\
&=& e^{ip\cdot a}\sum_{\chi'}D_{\chi'\chi}^j(W(\alpha, \Lambda^{-1}(\alpha)p))\phi_{\chi'}( \Lambda^{-1}(\alpha)p)
\end{eqnarray}
or equivalently 
\begin{equation}
U(\alpha, a)\kt{\wig} = e^{-i\Lambda(\alpha)p\cdot a}\sum_{\chi'}D_{\chi'\chi}^{j}(W(\alpha, p)) \kt{\Lambda(\alpha)p,\chi'[\ms,j]},
\end{equation}
\end{subequations}
where $D_{\chi'\chi}^j(u)$ is the $2j+1$ dimension representation of the quantum mechanical rotation $u\in{\rm SU}(2)$ and $W(\alpha, p)\in{\rm SU}(2)$ is the Wigner rotation.  For our choice of representant $\alpha(p)$,  $W(\alpha, p)$ has the form
\begin{equation}\label{wignerrot}
W(\alpha, p) = \ell^{-1}(\Lambda(\alpha)p)\alpha\ell(p),
\end{equation}
where $\ell(p)$ is the standard, rotation-free boost (see (\ref{ell})) satisfying $\Lambda(\ell(p))p_R = p$.

\section{Multiparticle bases and the CGCs of $\ptr$}

The general idea is that there are two bases appropriate for considering non-interacting, two-particle states and we want to write the transformation coefficients or CGC's between the two $\ptr$ representations, the direct product (which is not a UIR) and direct sum (which is).

The direct product basis has the form
\begin{equation}
\kt{1\otimes 2} = \kt{p_1 \chi_1 n_1; p_2 \chi_2 n_2} = \kt{p_1 \chi_1 n_1}\otimes\kt{p_2 \chi_2 n_2},
\end{equation}
where $n_i=\{\ms_i ,j_i\}$ characterizes the $\ptr$ UIR of particle $i$ and $\kt{p_i \chi_i n_i}\in\Phi^\times(\ms_i, j_i)=\Phi^\times(n_i)$.
We make the following choice of notation and normalization for the two-particle direct product states:
\begin{eqnarray}\label{normdp}
\bk{1 \otimes 2}{1' \otimes 2'} &=& \bk{p_1 \chi_1 n_1;p_2 \chi_2 n_2}{p'_1 \chi'_1 n_1;p'_2 \chi'_2 n_2}\nonumber\\
&=& 2E_1(\mmp_1) 2E_2(\mmp_2) \ms_1 \ms_2 \delta^3(\vp_1 - \vp_1')\delta^3(\vp_2 - \vp_2')\delta_{\chi_1 \chi'_1}\delta_{\chi_2 \chi'_2}.
\end{eqnarray}
These gives the expansion of a vector $\phi\in\Phi(n_1) \otimes \Phi(n_2)$ as
\begin{eqnarray}\label{expdp}
\phi &=& \int d\mu(1)\,d\mu(2) \kt{1\otimes 2}\bk{1\otimes 2}{\phi}\nonumber \\
&=& \frac{1}{\ms_1 \ms_2}\sum_{\chi_1,\chi_2}\int\frac{d^3\vp_1d^3\vp_2}{4 E_1(\mmp_1)E_2(\mmp_2)} \kt{p_1 \chi_1 n_1;p_2 \chi_2 n_2}\bk{p_1 \chi_1 n_1;p_2 \chi_2 n_2}{\phi}
\end{eqnarray}
On the direct product vectors, the representation of $\ptr$ on $\Phi(n_1) \otimes \Phi(n_2)$ (and its extension to $\Phi^\times(n_1) \otimes \Phi^\times(n_2)$) is the direct product of the one-particle transformation representations (\ref{wigtrans}):
\begin{equation}
U(\alpha,a)\kt{1\otimes 2} = U_1(\alpha,a)\kt{p_1 \chi_1 n_1}\otimes  U_2(\alpha,a)\kt{p_2 \chi_2 n_2}.
\end{equation}
The CSCO for this basis is the sum of the one-particles CSCOs
\[
\{M_1^2, W_1^2, {\bf P}_1, S_3(P_1)_1, M_2^2, W_2^2, {\bf P}_2, S_3(P_2)_2\},
\]
where, for example, the notation $M_1^2$ implies the natural extension to the direct product space, $M_1^2 \otimes I$.

Alternatively, we can choose an basis for the two-particle states such that each basis ket is an element of a UIR that appears in the irreducible decomposition of the direct product of two one-particle representations.  
The $\Phi(n_1) \otimes \Phi(n_2)$ representation is not irreducible but can be decomposed into a direct sum of UIRs:
\begin{equation}\label{bigoplus}
\Phi(n_1) \otimes \Phi(n_2)  = \bigoplus_{N, \eta}\Phi(N)_{\eta, \alpha_{12}},
\end{equation}
where $\alpha_{12} = \{n_1, n_2\}$.
The label $N$ specifies the transformation properties of the a single UIR $\Phi(N)_{\eta, \alpha_{12}}$ that occurs in the irreducible decomposition.  The direct sum over $N = \{\ms, j\}$ in the case of two massive UIRs is a direct integral over $\ms$ from the minimal center-of-mass-energy squared $(m_1 + m_2)^2$ to infinity and a direct sum over $\{j_0, j_0 +1, ...\}$, where $j_0=1/2$ if the UIRs correspond to a fermion and a boson or $j_0=0$ if they correspond to a like pair.  A particular value for $N$ may appear multiple times in the decomposition, and $\eta$ is a pair of numbers that labels this degeneracy.  In this paper, we will label the degeneracy according to the spin-orbit coupling scheme of Joos~\cite{joos} and MacFarlane~\cite{macfarlane}.  In this scheme, the single particle generators can be combined to form total intrinsic spin ${\bf S}$ and orbital angular momentum ${\bf L}$ operators (see \cite{kandpoly}, p.~329 for explicit constructions), and $\eta = \{s,l\}$ are their eigenvalues.  The label $\alpha_{12}$ is carried along because there are still invariants of the single particle generators.

So, choosing the spin-orbit scheme for coupling and using the representants $\ell(p)$, the CSCO of operators for the direct sum basis will be
\[
\{M^2, W^2, {\bf P}, S_3(P), {\bf L}^2, {\bf S}^2, M_1^2, W_1^2, M_2^2, W_2^2\},
\]
of which all but ${\bf P}$ and $S_3(P)$ are invariant under $\ptr$ transformations and so the basis vectors of different UIRs $\Phi(N)_{\eta, \al_{12}}$ are orthogonal.  We choose the normalization in the following fashion,
\begin{eqnarray}\label{normds}
\bk{\tau}{\tau'} &=& \bk{p \chi [\ms j l s\alpha_{12}]}{p' \chi' [\ms' j' l' s'\alpha_{12}]} \nonumber\\
&=& 2 E_\ms(\mmp)\ms^2 \delta^3(\vp - \vp')\delta_{\chi \chi'}\delta(\ms - \ms')\delta_{jj'}\delta_{ss'}\delta_{ll'},
\end{eqnarray}
which is consistent with our intention of a dimensionless CGC.
This gives an expansion of a vector $\phi\in\Phi(n_1) \otimes \Phi(n_2)$ (compare to (\ref{expdp}))
\begin{eqnarray}\label{expds}
\phi &=& \int d\mu(\tau) \kt{\tau}\bk{\tau}{\phi}\nonumber\\
&=& \sum_{j,l,s}\int \frac{d\ms}{\ms^2}\sum_\chi \int \frac{d^3\vp}{2 E_\ms(\mmp)} \kt{p \chi [\ms j l s\alpha_{12}]}\bk{p \chi [\ms j l s\alpha_{12}]}{\phi},
\end{eqnarray}
where the range for $\{l,s\}$ is determined from $\alpha_{12}$ as shown below.
These kets obey the same transformation rule (\ref{wigtrans}) as the one-particle UIRs:
\begin{equation}
U(\alpha, a)\kt{p \chi [\ms j l s\alpha_{12}]} = e^{-i\Lambda(\alpha)p\cdot a}\sum_{\chi'}D_{\chi'\chi}^{j}(W(\alpha, p)) \kt{\Lambda(\alpha)p \chi' [\ms j l s\alpha_{12}]}.
\end{equation}

The CGCs for the quantum mechanical Poincar\'e group then are the amplitudes
\[
\bk{1\otimes 2}{\tau} =\bk{p_1 \chi_1 p_2 \chi_2 \alpha_{12}}{p \chi [N \eta \alpha_{12}]}.
\]
Their structure clearly depends on how the one-particle UIRs are constructed (and therewith the one-particle CSCOs) as well as the choice of coupling scheme.  A general scheme for constructing the CGCs of $\ptr$ is the double-coset method~\cite{mackey} used in \cite{moussa,schaaf,klinksmith}.  Whippman~\cite{whippman} uses a nice alternative technique involving group integration over representation matrix elements.  We use spin-orbit coupling, the method of \cite{joos, macfarlane, kandpoly}, which exploits the fact that angular momentum coupling for two particles in their rest frame has the same form in relativistic and non-relativistic physics.  Alternatively, using the helicity basis (i.e., choosing the representant $\alpha(p)$ that leads to the helicity definition of the spin component), the CGCs are derived by Jacob and Wick~\cite{jacobwick} in a very intuitive fashion and these CGCs have been used extensively and extended to transversity~\cite{transversity}.

The CGCs for the direct product of two, distinguishable representations of $\ptr$ can be split into a kinematics/normalization term and an angular correlation term:
\begin{eqnarray}\label{cgcres}
\bk{p_1 \chi_1 p_2 \chi_2 \alpha_{12}}{p \chi [N \eta \alpha_{12}]} &=& K_{12}(p_1 p_2 ; p)A_{12}(p_1 \chi_1 p_2 \chi_2; p \chi j \eta)\nonumber\\
\mbox{or}\ \bk{1\otimes 2}{\tau} &=& K_{12}(1,2;\tau)A_{12}(1,2;\tau).
\end{eqnarray}

The term $K_{12}(1,2;\tau)$ is the kinematic term involving momentum conservation.  It depends on the normalizations (\ref{normdp}) and (\ref{normds}) and looks like:
\begin{equation}\label{kin}
K_{12}(p_1 p_2 ; p) = \left(\frac{64 \ms_1^2 \ms_2^2 \ms^2}{\lambda(\ms, \ms_1, \ms_2)}\right)^{1/4}    \ms\, 2E_\ms(\mmp)  \delta^3(\vp_1 + \vp_2 - \vp)\delta((p_1 + p_2)^2 - \ms),
\end{equation}
where
\[
\lambda(\ms, \ms_1, \ms_2) = \ms^2 + \ms_1^2 + \ms_2^2 - 2(\ms\ms_1 + \ms\ms_2 + \ms_1\ms_2).
\]
Note that the magnitude of the center-of-mass momentum of both particles is 
\[
\mk = \sqrt{\frac{\lambda(\ms, \ms_1, \ms_2)}{4\ms}}.
\]
A choice of phase convention has been made such that $K=K^*$.

The term $A_{12}(1,2;\tau)$ contains the information about the angular distribution and spin correlations.  For spin orbit-coupling,
\begin{eqnarray}\label{ang}
A_{12}(p_1 \chi_1 p_2 \chi_2; p \chi j \eta) &=& \sum_{\chi'_1 \chi'_2} D^{j_1}_{\chi'_1 \chi_1}(u(p, p_1)) D^{j_2}_{\chi'_2 \chi_2}(u(p, p_2))\nonumber\\
&&\times
\sum_{l_3 s_3} C(s j_1 j_2; s_3 \chi'_1 \chi'_2) C(j l s; \chi l_3 s_3)
(-)^\chi Y_{ll_3}(\hat{\bf \Omega}(p_1,p_2)),
\end{eqnarray}
where the following conventions and notations have been chosen:
\begin{itemize}
\item The rotation $W(\ell^{-1}(p),p_i)$ when applied to a single particle ket $\kt{p_i, \chi_i, n_i}$ effects the mixing of the spin coordinates when the basis vector for particle $i$ is transformed into barycentric frame in which the spin coupling takes place.  The argument $u(p, p_i)$ of the rotation matrix $D^{j_i}$ is then the inverse of this rotation:
\begin{equation}
u(p,p_i) = W^{-1}(\ell^{-1}(p),p_i).
\end{equation}

\item The CGC's for the rotation group are chosen according to standard phase conventions (i.e., they are all real).  They are
\begin{equation}
C(j j_1 j_2; \chi \chi_1 \chi_2) = \bk{\chi_1 \chi_2 [j_1 j_2]}{\chi [j j_1 j_2]}
\end{equation}
where $\chi = \chi_1 + \chi_2$.  These are used to couple the spins of the two particles and to couple the total spin with the orbital angular momentum.
\item The spherical harmonic $Y_{ll_3}(\hat{\bf \Omega})$ describes the angular dependence on the orbital angular momentum and is a function of the unit-normalized relative momentum $e= (0, \hat{\bf \Omega})$ in the barycentric frame:
\begin{equation}\label{sphe}
e(p_1,p_2) = \left(\frac{\ms}{\lambda(\ms,\ms_1,\ms_2)}\right)^{1/2} L^{-1}(p_1 + p_2)\{p_1 - p_2 - [(\ms_1 - \ms_2)/\ms](p_1 + p_2)\}.
\end{equation}
\item The phase factor $(-)^\chi$ is introduced for later convenience so that the direct sum basis vectors transform in the usual way under time reversal.  This could also have been effected by choosing the alternate convention for the spherical harmonics $Y_{ll_3}(\hat{\bf \Omega})\rightarrow (i)^lY_{ll_3}(\hat{\bf \Omega})$~\cite{edmunds}.
\end{itemize}

Necessary properties of $K(1,2,\tau)$ and $A(1,2,\tau)$ are established for later results in Appendix B.

\section{Comments on Internal Quantum Numbers, Superselection Sectors, and Symmetrization}

Multiparticle dynamics takes place in superselection sectors identified by a set of conserved, additive internal quantum numbers $q = \{q^1, q^2,...\}$ like baryon number, lepton number and charge.  Combining the representations of these symmetries with the UIRs of $\ptr$ is simple:  each UIR $\Phi(n)_q$ is now labeled by a set of numbers $q$ and all elements of $\Phi(n)_q$ are eigenstates of the set of operators $Q= \{Q^1, Q^2,...\}$.

In the direct product $\Phi(n_1)_{q_1}\otimes\Phi(n_2)_{q_2}$, since we are only concerned with Abelian internal symmetry groups, all UIRs in the direct sum will have the eigenvalues of $Q = Q_1 \otimes I + I \otimes Q_2$ as $q = \{q^1_1 + q_2^1, q_1^2 + q_2^2,...\}$, leading to the CGC for $\ptr$ and internal symmetries:
\begin{equation}
\bk{p_1 \chi_1 q_1 p_2 \chi_2 q_2 \alpha_{12}}{p \chi [N \eta q \alpha_{12}]} = K_{12}(p_1 p_2 ; p)A_{12}(p_1 \chi_1 p_2 \chi_2; p j \eta)\delta_{q,q_1 + q_2},
\end{equation}
although the inclusion of the delta function is unnecessary, since $Q_1$ and $Q_2$ are still elements of the two-particle CSCO.
A particularly interesting case, as will be discussed below, is when $q_1 = -q_2$ and therefore $q=0$ and the intrinsic charge parity has a physical meaning.

Additionally, another superselection rule prohibits superpositions of fermions on bosons.  The CGCs of $\ptr$ already satisfy this requirement; all angular coefficients $A_{12}$ (\ref{ang}) disappear if $j$ does not have the same integrality as $j_1 + j_2$.  In other words, the only non-zero CGCs for two fermions or two bosons have only integral $j$'s in the direct sum (\ref{bigoplus}); the direct product of a fermion and a boson result in only half-integral $j$'s.

The direct sum states above treated the particles as distinguishable, but now we must consider spin-statistics.  We choose a convention to deal with the interchange of particles that is consistent with the requirements of spin-statistics.  We make the choice, as in \cite{weinberg}, that a negative sign arise in the exchange of any two fermions, identical or not:
\begin{equation}\label{mpconv}
\kt{1 \otimes 2} = \left\{ \begin{array}{ll}
														-\kt{2 \otimes 1} & \mbox{if both fermions}\\
														\kt{2 \otimes 1} & \mbox{else}.\end{array}\right.
\end{equation}

\section{Adjoining discrete symmetries}

We now want to represent the physical symmetries $P$, $C$ and $T$ by unitary and antiunitary operators $U_P$, $U_C$ and $A_T$.  As with the restricted Poincar\'e group, finding the representations requires more than just adding the discrete symmetries to the algebra.  The symmetry transformations have to be modified to take into account that quantum mechanical probabilities require we deal with projective representations.  Also, we must incorporate the fact that some symmetry transformations must be represented by antiunitary operators if we want to maintain the positivity of the energy.  These two facts, as first shown by Wigner for $P$ and $T$ in~\cite{wigner64} and expanded to include $C$ by Goldberg~\cite{goldberg}, allow for 16 inequivalent UIRs for the full (quantum mechanical) Poincar\'e group $\pt$ with a given mass and spin.

We will choose the two conventional representations and the phases such that acting on the Wigner momentum basis of a one-particle UIR of $\pt$ we have:
\begin{eqnarray}\label{actioncpt}
U_P\kt{p_i\,\chi_i[\ms_i\,j_i\,\pi_i\,\xi_i\,q_i]} &=& \pi_i\kt{gp_i\,\chi_i[\ms_i\,j_i\,\pi_i\,\xi_i\,q_i]} = \pi\kt{P i}\nonumber\\
U_C\kt{p_i\,\chi_i[\ms_i\,j_i\,\pi_i\,\xi_i\,q_i]} &=& \xi_i\kt{p_i\,\chi_i[\ms_i\,j_i\,\pi_i\,\xi_i\,\bar{q}_i]}
= \xi_i\kt{Ci}\nonumber\\
A_T\kt{p_i\,\chi_i[\ms_i\,j_i\,\pi_i\,\xi_i\,q_i]} &=& (-)^{j-\chi}\kt{gp_i\,-\chi_i[\ms_i\,j_i\,\pi_i\,\xi_i\,q_i]} = (-)^{j-\chi}\kt{Ti},
\end{eqnarray}
where $g$ is the Minkowski metric with $g_{00}=1$ and $g_{ij}=-\delta_{ij}$ and can also be thought of as the $4\times 4$ representation of the parity operator $P$. The label $q$ stands for internal quantum numbers such a charge and $\bar{q}=-q$ are the internal quantum numbers of its antiparticle.  Note that we have now included explicitly in our notation for the basis ket two more quantum numbers, the intrinsic parity $\pi=\pm 1$ and intrinsic charge parity $\xi=\pm 1$.

The following relations further specify which of the 16 representations are chosen, one for massive bosons and one for massive fermions~\cite{goldberg,scurek}:
\begin{eqnarray}\label{repcho}
U_P^2 = U_C^2 = 1,&\ &A_T^2 = (-1)^{2j},\nonumber\\
(U_PA_T)^2 = (-1)^{2j},&\ &(U_CA_T)^2 = (-1)^{2j},\nonumber\\
(U_CU_P)^2 &=& \left\{ \begin{array}{ll}
											+1 & \mbox{for massive bosons}\\
											-1 & \mbox{for massive fermions}.
											\end{array}\right.
\end{eqnarray}
The last relation above reflects the fact that the parities of fermions and their antifermion partner are opposite.  The other 14 representations are mostly unused by current quantum field theory, although certain cases allow for time reversal doubling and may be useful for the incorporation of an irreversible time evolution semigroup into quantum theory~\cite{bohmsuj97}.
In these two representations, $\Phi(n)$ is no longer a UIR if the particle is charged, since $U_C$ transforms out of it, but $\Phi(n)\oplus\Phi(\bar{n})$ is a one-particle UIR of $\pt$, and $\Phi(n)\otimes\Phi(\bar{n})$ can be decomposed into UIRs of $\pt$, as discussed below.

\subsection{Parity}

We now have the machinery in place to extend the CGCs for the restricted Poincar\'e group with internal quantum numbers (\ref{cgcres}) to that of the full Poincar\'e group.  
  Beginning with parity, we consider the CGCs between direct product basis ket
\begin{equation}
\kt{p_1\,\chi_1 [\ms_1\,j_1\,\pi_1\,\xi_1\,q_1] p_2\,\chi_2 [\ms_2\,j_2\,\pi_2\,\xi_2\,q_2]} = \kt{1\otimes 2}\in\Phi^\times(n_1)_{q_1}\otimes\Phi^\times(n_2)_{q_2}
\end{equation}
and direct sum basis ket
\begin{equation}
\kt{p\,\chi [\ms\,j\, \pi\,\xi\,l\,s\, \alpha_{12}]}= \kt{\tau}\in\Phi^\times(N)_{\eta, \alpha_{12}, \pi, \xi},
\end{equation}
where now $\alpha_{12}=\{n_1 n_2 q_1 q_2\}$

The action of $U_P$ on the direct product kets is
\begin{eqnarray}
U_P\kt{p_1\,\chi_1 \,p_2 \,\chi_2 [\alpha_{12}]} &=& \pi_1\pi_2\kt{gp_1\,\chi_1 \,gp_2 \,\chi_2 [\alpha_{12}]}\nonumber \\
&=& \pi_1\pi_2\kt{P1\otimes P2}
\end{eqnarray}
and we define $\pi$ (the intrinsic parity of UIRs in the direct sum representation) by 
\begin{equation}
U_P\kt{p\,\chi [N\, \pi\,\xi\,\eta\, \alpha_{12}]} = \pi\kt{gp\,\chi [N\, \pi\,\xi\,\eta\, \alpha_{12}]}=\pi_\kt{P\tau}.
\end{equation}
Since $U_P=U_P^\dag$ and the $\pi$'s are real, we then have a relationship between the amplitudes
\begin{equation}\label{phop}
\pi_1\pi_2\bk{P1\otimes P2}{\tau} = \pi\bk{1\otimes 2}{P\tau}.
\end{equation}
If we can establish a relationship between $\bk{P1\otimes P2}{\tau}$ and $\bk{1\otimes 2}{P\tau}$, we will be able to ascertain the CGC coefficient including parity.

Using (\ref{pkin}), we find
\begin{equation}\label{kinp}
K_{12}(gp_1 gp_2 ; p) =K_{12}(p_1 p_2 ; gp) 
\end{equation}
and (\ref{angpb}) 
\begin{equation}
A_{12}(gp_1 \chi_1 gp_2 \chi_2; p \chi j \eta) = (-)^l A_{12}(p_1 \chi_1 p_2 \chi_2; gp \chi j \eta) 
\end{equation}
Putting this together, we have
\begin{equation}
	\bk{P1\otimes P2}{\tau} = (-)^l \bk{1\otimes 2}{P\tau}.
\end{equation}
Referring back to (\ref{phop}), then either the amplitude
\[
\bk{P1\otimes P2}{\tau} = \bk{1\otimes 2}{P\tau} = 0
\]
 or
\begin{equation}
\pi = \pi_1\pi_2 (-)^l.
\end{equation}
Phrasing this as a CGC, we have
\begin{equation}\label{cpcgcpta}
\bk{p_1\,\chi_1 \,p_2 \,\chi_2 [\alpha_{12}]}{p\,\chi [N\, \pi\,\xi\,\eta\, \alpha_{12}]} = K_{12}(p_1 p_2 ; p)A_{12}(p_1 \chi_1 p_2 \chi_2; p \chi j \eta) \delta_{\pi,\pi_1\pi_2 (-)^l}\Delta(\xi).
\end{equation}
where $\Delta(\xi)$ is a possible CGC for $\xi$ to be determined.  Note that this is not enough to check which representation of $\pt$ (e.g., whether $(U_C U_P)^2$ equals $+1$ or $-1$) the UIR $\Phi(N)_{\eta\, \alpha_{12}, \pi, \xi}$ is.  For that we also need to consider the actions of $U_C$ and $A_T$.

\subsection{Charge Parity}

The analysis of charge parity begins the same way; the action of $U_C$ on the basis kets is
\begin{eqnarray}
U_C\kt{p_1\,\chi_1 \,p_2 \,\chi_2 [\alpha_{12}]} &=& \xi_1\xi_2\kt{p_1\,\chi_1 \,p_2 \,\chi_2 [\bar{\alpha}_{12}]}\nonumber\\
&=& \xi_1\xi_2\kt{C1\otimes C2}
\end{eqnarray}
and
\begin{eqnarray}
U_C\kt{p\,\chi [N\, \pi\,\xi\,\eta\, \alpha_{12}]} &=& \xi\kt{p\,\chi [N\, \bar{\pi}\,\bar{\xi}\,\eta\, \bar{\alpha}_{12}]}\nonumber\\
&=& \xi\kt{C\tau}
\end{eqnarray}
where $\alpha_{12}=\{n_1, n_2, q_1, q_2\}$ and $\bar{\alpha}_{12}=\{n_1, n_2, -q_1, -q_2\}$.  As a consequence of the choice of representation (\ref{repcho}), particles and antiparticles have the same parity and charge parity if bosons but opposite parity (but same charge parity) if fermions.  Symbolically,
\begin{equation}
\xi_i = \bar{\xi}_i\ \mbox{and}\ \pi_i = \pm \bar{\pi}_i,
\end{equation}
with the minus sign for fermions.  Similarly, we know that all sixteen UIRs of $\pt$ satisfy $\xi = \bar{\xi}$, but we must check whether the relationship exists between $\pi$ and $\bar{\pi}=\bar{\pi}_1\bar{\pi}_2(-1)^l$ is consistent with (\ref{repcho}):
\begin{eqnarray}
	(U_C U_P)^2 \kt{p\,\chi [N\, \pi\,\xi\,\eta\, \alpha_{12}]} &= &
	 \pi_1 \pi_2 (-)^l U_C U_P U_C\kt{gp\,\chi [N\, \pi\,\xi\,\eta\, \alpha_{12}]} \nonumber \\
	&= &  \xi \pi_1 \pi_2 (-)^l U_C U_P \kt{gp\,\chi [N\, \bar{\pi}\,\xi\,\eta\, \bar{\alpha}_{12}]}\nonumber \\
	&=& \bar{\pi}_1 \pi_1 \bar{\pi}_2 \pi_2 \xi (-)^{2l} U_C  \kt{p\,\chi [N\, \bar{\pi}\,\xi\,\eta\, \bar{\alpha}_{12}]}\nonumber \\
	&=& \bar{\pi}_1 \pi_1 \bar{\pi}_2 \pi_2 \xi^2 \kt{p\,\chi [N\, \pi\,\xi\,\eta\, \alpha_{12}]}.
\end{eqnarray}
If both particles are fermions or bosons, the product $\bar{\pi}_1 \pi_1 \bar{\pi}_2 \pi_2 =1 $ and since the total spin $j$ will be integer in this case, the properties of the UIR $\Phi(N)_{\eta\, \alpha_{12}, \pi, \xi}$ are consistent with (\ref{repcho}).  If it is a mixed fermion-baryon pairing, then the product $\bar{\pi}_1 \pi_1 \bar{\pi}_2 \pi_2 = -1 $ and since $j$ will be odd-half-integer, it is still consistent with the UIR of $\pt$ selected by (\ref{repcho}).

This could be established because $\xi = \bar{\xi}$ in the UIRs of $\pt$, but what can be said about the value of $\xi$?  Since $U_C=U_C^\dag$ and the $\xi$'s are real by choice, we have a relationship between the amplitudes
\begin{equation}\label{chop}
\xi_1\xi_2\bk{C1\otimes C2}{\tau} = \xi\bk{1\otimes 2}{C\tau}.
\end{equation}
Unlike the case of parity, it is in general not possible to find a meaningful relationship between
$\bk{C1\otimes C2}{\tau}$ and $\bk{1\otimes 2}{C\tau}$.
It is only possible if $\alpha_{12} = \bar{\alpha}_{12}$, which allows for two cases: (1) the direct product particles are C-eigenstates ($q_i=0$), or (2) the direct product particles are charge conjugates ($q_1 = -q_2$).  For the other cases, $\xi$ has no absolute physical content since $\Phi(N)_{\eta\, \alpha_{12}, \pi, \xi}$ is not an eigenspace of $U_C$, and so $\xi$ may be redefined to either $\pm 1$ by the transformation $U_C \rightarrow \exp(i \theta Q)U_C$.

\begin{enumerate}
\item In the case $q_1 = q_2 = q = 0$,
\begin{eqnarray}
\kt{C1\otimes C2} &=& \xi_1 \xi_2 \kt{1\otimes 2}\ \mbox{and}\nonumber\\
\kt{C\tau} &=& \xi\kt{\tau},
\end{eqnarray}
and from (\ref{chop}) we have
\begin{equation}\label{npcgcpt}
\bk{p_1\,\chi_1 \,p_2 \,\chi_2 [\alpha_{12}]}{p\,\chi [N\, \pi\,\xi\,\eta\, \alpha_{12}]} = K_{12}(p_1 p_2 ; p)A_{12}(p_1 \chi_1 p_2 \chi_2; p \chi j \eta) \delta_{\pi,\pi_1\pi_2 (-)^l} \delta_{\xi,\xi_1\xi_2}.
\end{equation}

\item The other case $q=0$, $q_1 = -q_2$ is more interesting.  Using the notation $j_1 = j_2 = j_0$, $\ms_1 = \ms_2 = \ms_0$ (or $n_1 = n_2 = n_0$), $\xi_1=\xi_2=\xi_0 = \pm 1$ (although $\xi_0$ is arbitrary), and $\alpha_{21} = \{n_0, n_0, q_2, q_1\} = \bar{\alpha}_{12}$, we see that
\begin{eqnarray}
	U_C \kt{p_1\,\chi_1 \,p_2 \,\chi_2 [\alpha_{12}]} &=& \xi_0^2 \kt{p_1\,\chi_1 \,p_2 \,\chi_2 [\alpha_{21}]}\nonumber\\
	&=& (-)^{2j_0}\kt{p_2\,\chi_2 \,p_1 \,\chi_1 [\alpha_{12}]},
\end{eqnarray}
where the second equality follows from (\ref{mpconv}) and the fact that for fermions $(-)^{2j_0} = -1$ and for bosons $(-)^{2j_0} = 1$. Therefore,
\begin{eqnarray}
\bk{p_1\,\chi_1 \,p_2 \,\chi_2 [\alpha_{12}]}{p \chi [\ms j \pi\xi ls \alpha_{12}]}&=&
(-)^{2j_0}\bk{p_2\,\chi_2 \,p_1 \,\chi_1 [\alpha_{12}]}{p\,\chi [N\, \pi\,\xi\,\eta\, \alpha_{12}]}\nonumber \\
&=& (-)^{2j_0} K_{12}(p_2 p_1 ; p)A_{12}(p_2 \chi_2 p_1 \chi_1; p \chi l s).
\end{eqnarray}
From (\ref{ckin}) in Appendix B, we have
\begin{equation}
K_{12}(p_2 p_1 ; p) = K_{12}(p_1 p_2 ; p)
\end{equation}
and from (\ref{angc}), we have
\begin{equation}
	A_{12}(p_2 \chi_2 p_1 \chi_1; p \chi j \eta) = (-)^{l + s - 2j_0} A_{12}(p_1 \chi_1 p_2 \chi_2; p \chi j \eta).
\end{equation}

Combining this, we achieve the result
\begin{equation}
\bk{p_1\,\chi_1 \,p_2 \,\chi_2 [\bar{\alpha}_{12}]}{p\,\chi [N\, \pi\,\xi\,\eta\, \alpha_{12}]}= (-)^{l + s}\bk{p_1\,\chi_1 \,p_2 \,\chi_2 [\alpha_{12}]}{p\,\chi [N\, \bar{\pi}\,\xi\,\eta\, \bar{\alpha}_{12}]}
\end{equation}
and from (\ref{chop}) we have $\xi = (-)^{l +s}$ or expressing this as a CGC
\begin{equation}\label{paptcgc}
\bk{p_1\,\chi_1 \,p_2 \,\chi_2 [\alpha_{12}]}{p\,\chi [N\, \pi\,\xi\,\eta\, \alpha_{12}]} = K_{12}(p_1 p_2 ; p)A_{12}(p_1 \chi_1 p_2 \chi_2; p \chi j \eta) \delta_{\pi,\pi_1\pi_2 (-)^l} \delta_{\xi,(-)^{l + s}}
\end{equation}
and this agrees with results from two-body non-relativistic quantum mechanics for quarkonium, for example.

\end{enumerate}

This completes the results for the CGCs of $\pt$.  For coupling a particle-antiparticle pair, use (\ref{paptcgc}); for coupling two neutral (not just charge neutral, but all $q^i_1 = q^i_2 = 0$), use (\ref{npcgcpt}).  In the case where $q \neq 0$, then $\xi$ has no real significance and the use of (\ref{cpcgcpt}) is appropriate:
\begin{equation}\label{cpcgcpt}
\bk{p_1\,\chi_1 \,p_2 \,\chi_2 [\alpha_{12}]}{p\,\chi [N\, \pi\,\eta\, \alpha_{12}]} = K_{12}(p_1 p_2 ; p)A_{12}(p_1 \chi_1 p_2 \chi_2; p \chi j \eta) \delta_{\pi,\pi_1\pi_2 (-)^l}.
\end{equation}

There are two ways to think about these results: one can consider that the CGCs establish the values of $\pi$ and $\xi$ for the $\Phi(N)_{\eta\, \alpha_{12}, \pi, \xi}$ as function of the one-particle parities and charge parities and the coupling parameters $\eta$.  Alternatively, one can see that $\phi\in\Phi(N)_{\eta\, \alpha_{12}, \pi, \xi}$ implies angular correlations between the two particles which are manifest when it is re-expressed in the direct product basis.  This perspective has relevance in particular to studying decay products and will be explored in the example of $\Upsilon(4S)$ below.

\subsection{Time Reversal}

Finally, we turn to time reversal, which has a different character.  There is no intrinsic ``time-reversal parity'' quantum number, so we will satisfy ourselves by showing that the CGCs are consistent with the standard action of $A_T$ in (\ref{actioncpt}) and therefore that products of the two standard representations of $\pt$ are also a direct sum of standard representations

The action of $A_T$ of the direct product kets is
\begin{equation}
A_T\kt{p_1\,\chi_1 \,p_2 \,\chi_2 [\alpha_{12}]} = (-)^{j_1 + j_2 - \chi_1 - \chi_2}\kt{gp_1\,-\!\chi_1 \,gp_2 \,-\!\chi_2 [\alpha_{12}]} = -)^{j_1 + j_2 - \chi_1 - \chi_2} \kt{T1\otimes t2}
\end{equation}
and for consistency we would like to see that the action on the direct sum states is
\begin{equation}\label{atds}
A_T\kt{p\,\chi [N\, \pi\,\xi\,\eta\, \alpha_{12}]} = (-)^{j - \chi}\kt{gp\,-\!\chi [N\, \pi\,\xi\,\eta\, \alpha_{12}]}= (-)^{j - \chi}\kt{T\tau}.
\end{equation}

As a first step, consider the CGC $\bk{T1\otimes T2}{\tau}$.
In Appendix B, we have established (\ref{tkin}) that
\begin{equation}
K_{12}(gp_1 gp_2 ; p) = K_{12}(p_1 p_2 ; gp) = K_{12}^*(p_1 p_2 ; gp)
\end{equation}
and (\ref{angt})
\begin{eqnarray}
	A_{12}(gp_1 -\chi_1 gp_2 -\chi_2; p \chi j \eta)
&=& (-1)^{j_1 + j_2 - \chi_1 - \chi_2}(-1)^{j - \chi}A^*_{12}(p_1 \chi_1 p_2 \chi_2; gp -\chi l s).
\end{eqnarray}
So we have
\begin{equation}\label{tfull}
\bk{\tau}{T1\otimes T2} 
= (-)^{j_1 + j_2 - \chi_1 - \chi_2}(-)^{j - \chi}\bk{1\otimes 2}{T\tau}.
\end{equation}

Relation (\ref{tfull}) can be used show to that indeed (\ref{atds}) holds: 
\begin{eqnarray}
	A_T\kt{\tau} &=& A_T \int d\mu(1)d\mu(2) \bk{1\otimes 2}{\tau}\kt{1\otimes 2}  \nonumber\\
	&=& \int d\mu(1)d\mu(2) \bk{\tau}{1\otimes 2}(-)^{j_1 + j_2 - \chi_1 - \chi_2}\kt{T1\otimes T2}  \nonumber\\
&=&	\int d\mu(T1)d\mu(T2) \bk{\tau}{T1\otimes T2}(-)^{j_1 + j_2 + \chi_1 + \chi_2}\kt{1\otimes 2}\nonumber\\
&=&	\int d\mu(1)d\mu(2) \bk{\tau}{T1\otimes T2}(-)^{j_1 + j_2 - \chi_1 - \chi_2}\kt{1\otimes 2}\nonumber\\
&=&	\int d\mu(1)d\mu(2) \bk{1\otimes 2}{T\tau}(-)^{j - \chi}\kt{1\otimes 2}\nonumber\\
	&=& (-)^{j - \chi}\kt{gp\,-\!\chi [N\, \pi\,\xi\,\eta\, \alpha_{12}]},
\end{eqnarray}
where we have used $d\mu(T1) = d\mu(1)$, which is evident from (\ref{expdp}).

This establishes that the representation space $\Phi(N)_{\eta\, \alpha_{12}, \pi, \xi}$ fulfills (\ref{actioncpt}) and therewith the CGC coefficients for $\pt$ with internal additive quantum numbers are fully established and consistent with (\ref{repcho}).

\section{Example and Conclusion}

Let us consider the decay of a particle and calculate the angular correlations of the decay products assuming $CP$-conservation.  With the results above, partial wave analysis can be performed on scattering and decay processes that conserve $CP$ using the spin-orbit basis, which is irreducible with respect to $\pt$, including $U_C$.  Similar analyses can be performed in the helicity basis. However, the multiparticle helicity basis is not irreducible with respect to $\pt$; it can provide relations between scattering amplitudes.

As an example, we consider the $\Upsilon(4S)$, which has the following quantum numbers:
\begin{equation}\label{upnum}
j_\Upsilon^{\pi_\Upsilon\xi_\Upsilon} = 1^{- -} \qquad \ms_\Upsilon = M^2 = (10580\text{ MeV})^2  
\end{equation}
This particle is entirely neutral, i.e.\ $q=0$.
The $\Upsilon(4S)$ is a strongly-decaying resonance, observed as a peak in the $e^+ e^-$ cross-section with width-to-mass ratio $\Gamma/M \approx 10^{-3}$, which means that time scale of decay is a thousand times longer than the time scale of the energy/mass oscillation.

The primary decay channels of 
$\Upsilon(4S)$ are the particle-antiparticle pairs $B^+B^-$ and $B^0\bar{B}^0$.  The $B$'s have nearly identical kinematic quantum numbers
\begin{equation}
j_B^{\pi_B} = 0^{-} \qquad \ms_B = m^2 = (5280\text{ MeV})^2 
\end{equation}
but are charged (with non-electric charge quantum numbers in the base of the neutral $B$'s).  The $B$'s are observed as weakly-decaying states, with a width-to-mass ratio $\Gamma/M \approx 10^{-10}$.  Compared to the time scale of the $\Upsilon(4S)$ resonance, the $B$'s are effectively stable.

To analyze the decay correlations, we associate the $B$'s to UIRs of $\pt$.  This may seem inappropriate since now interactions must be included, however interaction-incorporating generators can be defined that still satisfy the defining algebraic relations of $\pt$ and an interacting CSCO $\{M^2, W^2, {\bf P}, S_3({\bf P}), U_P, U_C\}$ can be chosen~\cite{weinberg,kandpoly} even if the task of re-expressing these operators in terms of the interaction-free generators is difficult or impossible.

Because the decay transition matrix elements are invariant with respect to $\pt$, the strong interaction connects the state of the resonance $\Upsilon(4S)$ to a tower of UIRs in the direct sum decomposition of the decay products $\Phi(B)\otimes\Phi(\bar{B})$ which have a Breit-Wigner distribution of center-of-mass energies.  This construction can be made rigorous in the form of the relativistic Gamow vector~\cite{rgv}, but the details of this construction are not necessary to find the angular correlations of the decay products.  We will approximate the width of the $\Upsilon(4S)$ as negligible and select the single UIR connected to the resonance as $\Phi(\Upsilon)=\Phi(\ms_\Upsilon, j_\Upsilon)_{\eta_\Upsilon, \alpha_{B\bar{B}}, \pi_\Upsilon, \xi_\Upsilon}$. The value of allowed $\eta_\Upsilon$'s can be found from looking at (\ref{paptcgc}).  From the rotation group CGCs, $s=0$ and $l=1$ if the parity and charge parity delta functions are satisfied.

The basis ket $\kt{p_R \chi [\Upsilon(4S)]}\in\Phi^\times(\Upsilon)$ in the center-of-mass frame $p_R = (M, 0, 0, 0)$ with spin component $\chi$ can be expanded is the direct product basis to find the angular correlations implied for the decay products:
\begin{equation}
\kt{p_R \chi [\Upsilon(4S)]} = \frac{1}{m^4} \int\frac{d^3\vp_1d^3\vp_2}{4 E_m(\mmp_1)E_m(\mmp_2)} \kt{p_1 [B];p_2 [\bar{B}]}\bk{p_1 [B];p_2 [\bar{B}]}{p_R \chi [\Upsilon(4S)]}.
\end{equation}
The kinematic correlation term is
\begin{eqnarray}
K_{B\bar{B}}(p_1 p_2 ; p_R) &=& \left(\frac{64 m^8 M^4}{\lambda(M^2, m^2, m^2)}\right)^{1/4}    2 M^3  \delta^3(\vp_1 + \vp_2)\delta((p_1 + p_2)^2 - M^2)\nonumber\\
&=& \left(\frac{M}{\mk}\right)^{1/2} \frac{4m^2 M^3}{\mmp_1^2}\delta(\mmp_1 - \mmp_2) \delta^2(\hat{\bf \Omega}_1 + \hat{\bf \Omega}_2)\delta(4E_m(\mmp_1)^2 - M^2)\nonumber\\
&=& \frac{m^2}{2}\left(\frac{M}{\mk}\right)^{7/2}\delta(\mmp_1 - \mk)\delta(\mmp_2 - \mk)\delta^2(\hat{\bf \Omega}_1 + \hat{\bf \Omega}_2).
\end{eqnarray}
The angular term is particularly simple since the $B$'s are spinless.
\begin{eqnarray}
A_{B\bar{B}}(p_1 p_2 ; p_R \chi j l s) &=& A_{B\bar{B}}(p_1 p_2 ; p_R \chi 1 1 1)\nonumber\\
&=& \sum_{l_3 s_3} C(0 0 0; s_3 0 0) C(1 1 0; \chi l_3 s_3)
(-)^\chi Y_{1l_3}(\hat{\bf \Omega}(p_1,p_2))\nonumber\\
&=& (-)^\chi Y_{1\chi}(\hat{\bf \Omega}_1).
\end{eqnarray}
Putting this together, we get
\begin{eqnarray}\label{upstate}
\kt{p_R \chi [\Upsilon(4S)]} &=&  \frac{1}{m^4} \int \mmp_1^2 \mmp_2^2 \frac{d\mmp_1d\mmp_2d^2\hat{\bf \Omega}_1d^2\hat{\bf \Omega}_2}{4 E_m(\mmp_1)E_m(\mmp_2)} \kt{p_1 [B];p_2 [\bar{B}]}(-)^\chi Y_{1\chi}(\hat{\bf \Omega}_1)\nonumber\\
&& \times \frac{m^2}{2}\left(\frac{M}{\mk}\right)^{7/2}\delta(\mmp_1 - \mk)\delta(\mmp_2 - \mk)\delta^2(\hat{\bf \Omega}_1 + \hat{\bf \Omega}_2)\nonumber\\
&=&  (-)^\chi\frac{(M^3\mk)^{1/2}}{2m^2} \int  d^2\hat{\bf \Omega}_1  Y_{1\chi}(\hat{\bf \Omega}_1) \kt{(\mk\hat{\bf \Omega}_1) [B];(-\mk\hat{\bf \Omega}_1) [\bar{B}]}.
\end{eqnarray}
This formula for the rest frame can then be boosted and/or rotated into any other reference frame (e.g., the lab frame) using the direct product representation of $\ptr$ (\ref{wigtrans}). 

Let us now specialize to the case of $\Upsilon(4S)$'s produced at positron-electron B-factories.  The $\Upsilon$'s are produced via annihilation into a photon that decays into a $b\bar{b}$ bound quark pair; as a result the $\Upsilon(4S)$ are primarily produced in superpositions (or mixtures) of the $\chi=\pm 1$ states.  Restricting $0 \leq \theta \leq \pi/2$, calling $k = (E_m(\mk), \mk\hat{\bf \Omega})$ and $gk = (E_m(\mk), -\mk\hat{\bf \Omega})$, and using the explicit form for the sphereical harmonics, equation (\ref{upstate}) can be rewritten as
\begin{eqnarray}
\kt{p_R \chi=\!\pm\! 1 [\Upsilon(4S)]} &=& \pm \frac{(M^3\mk)^{1/2}}{2m^2} \int  d^2\hat{\bf \Omega} \sin(\theta)e^{\pm i \phi}\nonumber\\
&& \times \left\{ \kt{k [B];gk [\bar{B}]} - \kt{gk [B];k [\bar{B}]}\right\},
\end{eqnarray}
which agrees, up to normalization and phase, with the starting point of the two-time formalism analysis of CP-violation in coherent $B^0\bar{B}^0$ state by the BABAR Collaboration (see equation 1.36 in~\cite{babar}).  At this point it is prudent to stress that CGC results are purely kinematical, deriving entirely from conservation laws.  Dynamical assumptions have to be made for applications, as in the case above where the dynamics explains why the $\chi=0$ component is suppressed.

CGCs for $\ptr$ and $\pt$ have many applications and much work is still to be done on the subject. Some results for CGCs of multi-particle ($N>2$) UIRs have been explored~\cite{kummer,klinksmith}, but these have not been extended to $\pt$ including C.  Also, for some applications, this analysis must been carried out for other classes of UIRs of $\pt$, including the massless and space-like representations of $\ptr$, as well as the non-standard UIRs of $\pt$ with time-reversal doubling.  A particular application of note in which the CGCs are necessary and in which work is already underway is in the use of CGCs to understand the scattering boundary conditions and irreversibility inherent to resonance processes within the Hardy-class space hypothesis~\cite{bohmPRA}.

\begin{acknowledgments}
The authors would like to thank the Physics and Astronomy Department of Rice University for their support during the beginning of this work.  One of us (NLH) would like to thank the NSF Young Researchers Travel Grant program and American University for supporting this research and its presentation at conferences.
\end{acknowledgments}

\appendix

\section{UIRs of $\ptr$}
Representations of $\ptr$ are constructed using the method of induced representations, first carried out by Wigner~\cite{wigner39} and generalized by Mackey~\cite{mackey}.  The technique relies on building the representation for the full group $\ptr$ from representations of a subgroup, for massive representations typically chosen to be $H = \su \times \mathbb{R}^4$ (see \cite{ruhl, schaaf} for details).

The $\tilde{\mathcal{P}}^\uparrow_+$ is a 10-parameter Lie group whose generators fulfill the following commutation relations~\cite{weinberg}:
\begin{equation*}
 [J_i,J_j] = i\epsilon_{ijk}J_k \ \ [J_i,K_j] = i\epsilon_{ijk}K_k
\end{equation*}
\begin{equation*}
 [K_i,K_j] = -i\epsilon_{ijk}J_k \ \ [J_i,P_j] = i\epsilon_{ijk}P_k
\end{equation*}
\begin{equation*}
 [K_i,P_j] = -i\delta_{ij}H \  \ [K_i,H] = -iP_i
\end{equation*}
\begin{equation}\label{Palg2}
 [J_i,H] = [P_i, H] = [P_i, P_j] = 0.
\end{equation}
The unitary representation of $\tilde{\mathcal{P}}^\uparrow_+$ are are direct sum of unitary irreducible representations (UIRs) on which the Casimir invariants of the algebra (\ref{Palg2}) act as multiples of the identity.  The two invariant operators are identified as the mass-squared $M^2 = P_\mu P^\mu$ and the negative square of the Pauli-Lubanski vector $W^2 = -w_\mu w^\mu$, where the four vector $w$ is
\begin{equation}
w_\mu = ({\bf P}\cdot{\bf J}, H{\bf J} + {\bf P}\times{\bf K})
\end{equation}
and $P^\mu = (H, {\bf P})$.
We will consider UIRs with positive definite mass labeled by $\ms =m^2$ and intrinsic spin $j$ such that for all $\phi\in\Phi(\ms,j)$,
\begin{equation}
M^2\phi = \ms\phi\ \mbox{and}\ W^2\phi = \ms j(j+1)\phi.
\end{equation}
The representation space $\Phi(\ms,j)$ is typically endowed with a topology given by the scalar product norm and is therefore the Hilbert space $\mathcal{H}(\ms,j)$; however, we instead will work with a dense subspace $\Phi(\ms,j)\subset\mathcal{H}(\ms,j)$ with a stronger topology that allows for the nuclear spectral theorem and the continuity of the Poincar\'e algebra (for the general details of rigged Hilbert spaces or Gel'fand triplets, see \cite{bohmrhs}; for details relevant to the Poincar\'e algebra see \cite{rgv}).   

Within a particular UIR, a complete set of commuting operators is chosen of the form
\[
\{M^2, W^2, {\bf P}, \Sigma_3(P)\},
\]
where the operator $\Sigma_3(P)$ is a function of the 4-vector momentum operator $P$ and is constructed as
\begin{equation}\label{spinop}
\Sigma_\mu(\vP) = \sqrt{\ms}^{-1}U(\alpha(P))w_\mu U^{-1}(\alpha(P))= \sqrt{\ms}^{-1}\Lambda(\alpha(P))w_\mu.
\end{equation}
The operator $U(\alpha(P))$ (and its $4\times 4$ representation $\Lambda(\alpha(P))$ is an operator associated to a particular ``boost'' element $\alpha(p)$.  Since $[P_\mu, w_\nu]=0$, $P$ can be replaced with its eigenvalue $p$ when $\Sigma_3(P)$ acting on a momentum eigenvector.
The group element $\alpha(p)$ has the property that it boosts the four momentum in the rest frame $p_R = (m, {\bf 0})$ to final momentum $p$, i.e.
\begin{equation}\label{boost}
\Lambda(\alpha(p))p_R = p.
\end{equation}
The choice of $\alpha(p)$ is not unique.  For $u\in\su$, the group
element $\alpha(p)u$ also fulfills (\ref{boost}).  The 4-momentum hyperboloid $p^2=\ms$ is isomorphic to the left coset space $Q = \slg/\su$ and the particular left coset $Q(p)=\{\alpha(p)\}$ contains all elements that satisfy (\ref{boost}).

Specifying which representative element $\alpha(p)$ of $Q(p)$ to use in (\ref{spinop}) gives different physical meanings for the spin component~\cite{kandpoly,kummer}.  Choosing $\alpha(p)$ as $\ell(p)$, defined as
\begin{equation}\label{ell}
\ell(p) = {\left( \frac{\sigma^\mu p_\mu}{m} \right)}^{1/2} = \frac{\hat{m} + \sigma^\mu p_\mu}{[2m(m+ E_\ms(\mmp))]^{-1/2}},
\end{equation}
(where $\sigma^\mu = (1_2, {\bf \sigma})$, $m=\sqrt{\ms}$, $\hat{m} = m 1_2$, $p = (E_\ms(\mmp), \vp)$, $E_\ms(\mmp) = \sqrt{\ms + \mmp^2}$, and $\vp^2 = \mmp^2$) means the realization $\Lambda(\ell(p))=L(p)$ is the standard, rotation-free boost.  Then we call $\Sigma_i(P)=S_i(P)$, and physically it is the $i$-th spin component in the particle rest frame.

Choosing $\alpha(p)=\rho(p)\ell(p_z)$, where $p_z = (E_\ms(\mmp), 0, 0, \mmp)$ and $\rho(p)\in\su$ is the rotation that takes the 3-axis into the direction of $p$
 makes the spin operator into the helicity operator:
\begin{equation}
\Sigma_3(p) = H(p) = \frac{{\bf J}\cdot {\bf p}}{|{\bf p}|}.
\end{equation}
Many other choices for a representant $\alpha(p)\in Q(p)$ are possible, another notable example being the choice that leads to `transversity'~\cite{transversity}.  In this case the $\al(p)$ is chosen with a different initial rotation so that it defines the spin component in the direction perpendicular to the plane in which subsequent decays take place.

The choice $\{M^2, J^2, P_i, S_3(p)\}$ leads to the Wigner 3-momentum spin basis for the expansion of the $\Phi(m, j)$ and the basis $\{M^2, J^2, P_i, H(p)\}$ leads to the helicity basis.

\section{Properties of $A_{12}$ and $K_{12}$}

Here we summarize useful properties of $K_{12}(1,2,\tau)$ and $A_{12}(1,2,\tau)$. We will use the following shorthand notations:
\begin{eqnarray}
P1 = \{gp_1, \chi_1, n_1\}\quad &P2 = \{gp_2, \chi_2, n_2\}&\quad P\tau = \{gp, \chi N, \eta\}\nonumber\\
T1 = \{Tp_1, -\chi_1, n_1\}\quad &
T2 = \{Tp_2, -\chi_2, n_2\}&\quad
T\tau = \{gp, -\chi, N, \eta\},
\end{eqnarray}
where $g_{00} = 1$ and $g_{ij} = -\delta_{ij}$.

Inspecting (\ref{kin}), we find
\begin{eqnarray}\label{pkin}
K_{12}(P1,P2,\tau) &=& \left(\frac{64 \ms_1^2 \ms_2^2 \ms^2}{\lambda(\ms, \ms_1^2, \ms_2^2)}\right)^{1/4}  2  \ms\, E_\ms(\mmp)  \delta^3(-\vp_1 - \vp_2 - \vp)\delta((gp_1 + gp_2)^2 - \ms)\nonumber\\
&=& \left(\frac{64 \ms_1^2 \ms_2^2 \ms^2}{\lambda(\ms, \ms_1^2, \ms_2^2)}\right)^{1/4}  2  \ms\, E_\ms(\mmp)  \delta^3(\vp_1 + \vp_2 + \vp)\delta((p_1 + p_2)^2 - \ms)\nonumber\\
&=& K_{12}(1,2,P\tau) 
\end{eqnarray}
and similarly 
\begin{equation}\label{tkin}
K_{12}(T1,T2,\tau) = K_{12}(1,2,T\tau).
\end{equation}
Also we have,
\begin{equation}\label{ckin}
K_{12}(2,1,\tau) = K_{12}(1,2,\tau).
\end{equation}

Inspecting (\ref{ang}), we find
\begin{eqnarray}\label{angpa}
	A_{12}(P1,P2,\tau) &=& \sum_{\chi'_1 \chi'_2} D^{j_1}_{\chi'_1 \chi_1}(u(p,gp_1)) D^{j_2}_{\chi'_2 \chi_2}(u(p,gp_2)))\\
&&\times
\sum_{l_3 s_3} C(s j_1 j_2; s_3 \chi'_1 \chi'_2) C(j l s; \chi l_3 s_3)
(-)^\chi Y_{ll_3}(\hat{\bf \Omega}(gp_1,gp_2))
\end{eqnarray}
Using the properties that  $L(gp)=gL(p)g$ and $g\La(\rho)g$ for $\rho\in\su$, we find
\begin{eqnarray}\label{rotp}
W(\ell^{-1}(p), gp_i) &=& W(\ell^{-1}(gp),p_i)\ \mbox{and so }\
u(p,gp_i) = u (gp, p_i),
\end{eqnarray}
which also could have been derived from the fact that $[U_P, {\bf J}]=0$.  Also,
from (\ref{sphe}), we find that
\begin{equation}\label{omp}
\hat{\bf \Omega}(gp_1,gp_2) = - \hat{\bf \Omega}(p_1,p_2).
\end{equation}
Then using $Y_{ll_3}(-\hat{\bf \Omega})=(-)^lY_{ll_3}(\hat{\bf \Omega})$, we have 
\begin{eqnarray}\label{angpb}
	A_{12}(P1,P2,\tau) &=& (-1)^l \sum_{\chi'_1 \chi'_2} D^{j_1}_{\chi'_1 \chi_1}(u (gp, p_1)) D^{j_2}_{\chi'_2 \chi_2}(u (gp, p_2))\\
&&\times
\sum_{l_3 s_3} C(s j_1 j_2; s_3 \chi'_1 \chi'_2) C(j l s; \chi l_3 s_3)
(-)^\chi Y_{ll_3}(\hat{\bf \Omega}(p_1,p_2))\nonumber\\
&=& (-)^l A_{12}(1,2,P\tau) 
\end{eqnarray}

Also from (\ref{ang}), we find
\begin{eqnarray}\label{angca}
	A_{12}(2,1,\tau) &=& \sum_{\chi'_1 \chi'_2} D^{j_0}_{\chi'_1 \chi_2}(u(p, p_2)) D^{j_0}_{\chi'_2 \chi_1}(u(p, p_1))\nonumber\\
&&\times
\sum_{l_3 s_3} C(s j_1 j_2; s_3 \chi'_1 \chi'_2) C(j l s; \chi l_3 s_3)
(-)^\chi  Y_{ll_3}(\hat{\bf \Omega}(p_2,p_1))\nonumber\\
&=& \sum_{\chi'_1 \chi'_2} D^{j_0}_{\chi'_2 \chi_1}(u(p, p_1)) D^{j_0}_{\chi'_1 \chi_2}(u(p, p_2))\\
&&\times
\sum_{l_3 s_3} C(s j_0 j_0; s_3 \chi'_2 \chi'_1) C(j l s; \chi l_3 s_3)
(-)^\chi Y_{ll_3}(\hat{\bf \Omega}(p_2,p_1)),\nonumber
\end{eqnarray}
where the dummy indices in the second equality have been relabeled $\chi'_1 \rightarrow \chi'_2$, $\chi'_2 \rightarrow \chi'_1$.  Then, using~\cite{rose}
\[
C(s j_1 j_2; s_3 \chi'_2 \chi'_1) = (-)^{s - j_1 - j_2}C(s j_1 j_2; s_3 \chi'_1 \chi'_2),
\]
and $\hat{\bf \Omega}(p_2,p_1)= -\hat{\bf \Omega}(p_1,p_2)$, the relation (\ref{angca}) becomes
\begin{eqnarray}\label{angc}
	A_{12}(2,1,\tau) &=& (-)^{l + s - 2j_0} \sum_{\chi'_1 \chi'_2} D^{j_0}_{\chi'_1 \chi_2}(u(p, p_2)) D^{j_0}_{\chi'_2 \chi_1}(u(p, p_1))\nonumber\\
&&\times
\sum_{l_3 s_3} C(s j_0 j_0; s_3 \chi'_1 \chi'_2) C(j l s; \chi l_3 s_3)
(-)^\chi Y_{ll_3}(\hat{\bf \Omega}(p_1,p_2))\nonumber\\
&=& (-)^{l + s - 2j_0} A_{12}(1,2,\tau).
\end{eqnarray}

Finally to address
\begin{eqnarray}\label{angta}
	A_{12}(T1\otimes T2, \tau) 
	&=&  \sum_{\chi'_1 \chi'_2} D^{j_1}_{\chi'_1 -\chi_1}(u(p, gp_1)) D^{j_2}_{\chi'_2 -\chi_2}(u(p, gp_2)) \\
&&\times
\sum_{l_3 s_3} C(s j_1 j_2; s_3 \chi'_1 \chi'_2) C(j l s; \chi l_3 s_3)
(-)^\chi  Y_{ll_3}(\hat{\bf \Omega}(gp_1,gp_2)),\nonumber
\end{eqnarray}
we need the results (\ref{rotp}), (\ref{omp}), and the relation~\cite{rose}
\begin{equation}
(D^j_{m'm}(u))^* = (-)^{m'-m}D^j_{-m' -m}(u).
\end{equation}
Then (\ref{angta}) becomes
\begin{eqnarray}\label{angtb}
	A_{12}(T1\otimes T2, \tau) &=&  \sum_{\chi'_1 \chi'_2} (-)^{-\chi'_1 - \chi_1 -\chi'_2 - \chi_2}(D^{j_1}_{-\chi'_1 \chi_1}(u(gp, p_1)))^* (D^{j_2}_{-\chi'_2 \chi_2}(u(gp, p_2)))^* \nonumber\\
&&\times
\sum_{l_3 s_3} C(s j_1 j_2; s_3 \chi'_1 \chi'_2) C(j l s; \chi l_3 s_3)
(-)^{\chi + l} Y_{ll_3}(\hat{\bf \Omega}(p_1,p_2)).
\end{eqnarray}
We now relabel all the dummy indices by their negative, i.e.\ $\chi'_1 \rightarrow -\chi'_1$, $s_3 \rightarrow -s_3$, etc., and use the relation~\cite{rose}
\[
C(j j_1 j_2; m m_1 m_2) = (-)^{j_1 + j_2 - j}C(j j_1 j_2; -m -m_1 -m_2).
\]
and (\ref{angtb}) becomes
\begin{eqnarray}\label{angtc}
	A_{12}(T1\otimes T2, \tau) &=&  \sum_{\chi'_1 \chi'_2} (-1)^{\chi'_1 - \chi_1 + \chi'_2 - \chi_2}(D^{j_1}_{\chi'_1 \chi_1}(u(gp, p_1)))^* (D^{j_2}_{\chi'_2 \chi_2}(u(gp, p_2)))^*\nonumber\\
&&\times
\sum_{l_3 s_3} (-)^{j_1 + j_2 - j}C(s j_1 j_2; s_3 \chi'_1 \chi'_2) C(j l s; -\chi l_3 s_3)\nonumber\\
&&\times
(-)^\chi  Y_{l -l_3}(\hat{\bf \Omega}(p_1,p_2)).
\end{eqnarray}
Because of the presence of the rotation group CGCs, we have $l_3 + s_3 = -\chi$ and $\chi'_1 + \chi'_2 = s_3$.  Also, $Y_{l -l_3}(\hat{\bf \Omega}) = (-)^{l_3}Y^*_{l l_3}(\hat{\bf \Omega})$.  Inserting these relations and simplifying, (\ref{angtc}) becomes
\begin{eqnarray}\label{angt}
	A_{12}(T1\otimes T2, \tau) &=&  (-)^{j_1 + j_2 - \chi_1 - \chi_2 -j + \chi}\sum_{\chi'_1 \chi'_2} (D^{j_1}_{\chi'_1 \chi_1}(u(gp, p_1)))^* (D^{j_2}_{\chi'_2 \chi_2}(u(gp, p_2)))^* \nonumber\\
&&\times
\sum_{l_3 s_3} C(s j_1 j_2; s_3 \chi'_1 \chi'_2) C(j l s; -\chi l_3 s_3)
(-1)^{-\chi}   Y^*_{l -l_3}(\hat{\bf \Omega}(p_1,p_2))\nonumber\\
&=& (-1)^{j_1 + j_2 - \chi_1 - \chi_2}(-1)^{j - \chi}A^*_{12}(1\otimes 2, T\tau),
\end{eqnarray}
where we have used the facts that $((-)^\chi)^* = (-)^{-\chi}$ holds for integer and half-integer $\chi$ and $(-)^{-j + \chi} = (-)^{j - \chi}$ holds because $j - \chi$ is an integer.

\end{document}